\documentclass{article}
\usepackage{authblk}
\usepackage{geometry}
\usepackage{soul}
\usepackage{cite}
\usepackage{lipsum}
\usepackage[utf8]{inputenc}
\usepackage{graphicx}
\usepackage{epstopdf}
\usepackage{algorithmic}
\ifpdf
  \DeclareGraphicsExtensions{.eps,.pdf,.png,.jpg}
\else
  \DeclareGraphicsExtensions{.eps}
\fi
\usepackage[caption=false]{subfig}
\usepackage{amsmath,amssymb,amsfonts}
\usepackage{verbatim}
\usepackage{xcolor}
\usepackage{siunitx}
\newcommand{\ket}[1]{|#1\rangle}

\title{Nitrogen-Vacancy Emission from Nanodiamond: Size, Depth, and Surroundings}

\author[1,2]{Harini Hapuarachchi}
\author[1,2]{Francesco Campaioli}
\author[1,2]{Jared H Cole}
\author[1,2,3]{Andrew D Greentree}
\author[1,2,3,*]{Qiang Sun}
\affil[1]{Department of Physics, School of Science, RMIT University, Melbourne, VIC, 3001, Australia}
\affil[2]{RMIT Applied Quantum Technologies Centre, RMIT University, Melbourne, VIC, 3001, Australia}
\affil[3]{Australian Research Council Centre of Excellence for Nanoscale BioPhotonics, RMIT University, Melbourne, VIC, 3001, Australia}
\affil[*]{Corresponding author: qiang.sun@rmit.edu.au}

\date{}

\begin{document}

\maketitle

\begin{abstract}

    The negatively charged nitrogen-vacancy (NV) center in diamond is a leading solid-state quantum emitter, offering spin-photon interfaces over a wide temperature range with applications from electromagnetic sensing to bioimaging. While NV centers in bulk diamond are well understood, embedding them in nanodiamond (ND) introduces complexities from size, NV location, and NV polarizations. NVs in ND show altered fluorescence properties including longer lifetimes, lower quantum efficiency, and higher sensitivity to dielectric surroundings, which arise from radiative suppression, surface-induced non-radiative decay, and escape inefficiency at the diamond-background interface. Prior models typically addressed isolated aspects, such as dielectric contrast or surface quenching, without integrating full quantum-optical NV behavior with classical electrodynamics. We present a hybrid framework coupling rigorous electromagnetic simulations with a quantum-optical NV model including phonon sideband dynamics. NV emission is found to depend strongly on ND size, NV position, and surrounding refractive index. Our results explain observations such as shallow NVs in water-coated ND appearing brighter than deeper ones in air. This integrated model provides a unified framework for realistic NV in ND emission scenarios and informs the design of efficient NV-based sensors and quantum devices, advancing understanding of quantum emitter photophysics in nanoscale crystals.
    
\end{abstract}


\section{Introduction}
The negatively charged nitrogen-vacancy (NV) color centre in diamond has emerged as a premier platform for quantum science and technology~\cite{Doherty2013-PR}. The unique combination of properties of  NV centres, an optically addressable spin with long coherence times across a wide temperature range, and bright, stable single-photon emission~\cite{Doherty2013-PR, aharonovich2011diamond}, underpins a host of applications in quantum information~\cite{Childress2013-MRSB, Xu2019-OME}, metrology~\cite{Pham2012-PRB, Chen2017-NSR, Chigusa2025-PRD, Basu2025-ACSOmega}, and nanoscale sensing and imaging~\cite{Hong2013-MRSB, Schirhagl2014-ARPC, Doherty2014-PRL, Acosta2013-MRSB, Jeske2016-NJP, Bian2021-NatComm, Rovny2022-Science, Segawa2023-PNMRS, Paudel2024-Nanomaterials, Katsumi2025-CommMat}. These attributes have motivated intense research across disciplines: NV centres are being explored for quantum networks and computing, high-resolution bio-imaging, and precision sensing in physics and biology.

While initial NV experiments were performed in high-purity bulk diamond, many emerging applications rely on nanodiamond (ND) hosting NV centres. Certain use-cases demand NVs in nanoscale diamonds that can be dispersed, manipulated, or integrated into devices where bulk crystals are impractical~\cite{Plakhotnik2018-DRM}. However, the transition from bulk to nanoscale introduces challenges, in particular the inhomogeneity of NV emissions in ND: NVs in ND suffer from reduced quantum efficiency, broader emission lines, and spectral instability even when obtained from the same batch or nominally identical fabrication processes~\cite{Inam2013-APL}. These limitations depend strongly on nanodiamond size, NV position, and surrounding refractive index, making it critical to understand and control NV photophysics at the nanoscale for both application development and fundamental studies of light-matter interactions~\cite{Plakhotnik2018-DRM}.

Extensive research has shown that NV centres in ND exhibit markedly different optical behavior from those in bulk diamond due to both internal and external effects~\cite{Tisler2009-ACSNano, Inam2013-APL, Khalid2015-SR, Christinck2020-APB}. Early studies revealed that NV fluorescence lifetimes could increase in smaller NDs, which might be attributed to suppressed radiative rates due to altered electromagnetic environments. However, this suppression also competes with enhanced non-radiative pathways~\cite{Tisler2009-ACSNano, Inam2013-APL}. The quantum efficiency (QE) of NVs has been found to decrease significantly in smaller NDs~\cite{Berthel2015-PRB}. Systematic studies highlight that NV position within the ND plays a central role in decay rate modification and photostability~\cite{Radtke2019-NF, Sun2023-PRA}. In particular, when NV is close to the ND surface, nonradiative decay channels arising from sources such as band bending effects, surface adsorbates, chemical terminations, or structural defects may affect the decay rate modification~\cite{Hauf2011-PRB, Hu2013-CTC, Karaveli2016-PNAS}. 

% \SQ{In practical systems, additional nonradiative decay channels may be present, often arising from sources such as surface adsorbates, chemical terminations, or structural defects. These channels could potentially act as absorbers or perturb the local charge environment around the NV centre, leading to emission quenching. Since the specifics of these effects can vary significantly from case to case, they are not incorporated into our model. Instead, we focus on the computationally tractable general case of radiative decay, allowing us to isolate and quantify how the electromagnetic environment, specifically the dielectric structure surrounding the NV centre, modifies its emission properties.}

% \SQ{In addition to these surface-induced excitation reduction effects, band bending effects near the nanodiamond surface can also reduce the NV center’s expected average brightness. This band bending can shift the NV charge-state equilibrium or alter the internal field environment, affecting excitation efficiency or emission rates~\cite{Hauf2011-PRB, Hu2013-CTC}. While our model does not account for band bending, the emission suppression we observe due to radiative and non-radiative pathways is expected to occur alongside these electronic effects in real systems.}

From an optics perspective, two mechanisms must be distinguished. First, the high refractive index of diamond suppresses radiative emission by reducing the local density of optical states (LDOS) available to the NV dipole~\cite{Purcell1946-PR}. Second, the refractive index mismatch at the diamond-background interface leads to poor light outcoupling and reduced radiative rates\cite{Chew1988-PRA}. For example, experimental efforts show that embedding NDs in higher-index environments (e.g., water, PMMA) enhances emission rate and fluorescence brightness~\cite{Khalid2015-SR}.  
% Both effects contribute to the modification of NV brightness in nanodiamond and are treated explicitly in our framework. 

NV emission patterns and directionality depend on both dipole orientation and NV location within the ND, with central vs. off-centre placement significantly affecting NV emission profiles~\cite{Christinck2020-APB}. To enhance emission, researchers have employed coupling to photonic and plasmonic structures. NVs interfaced with metasurfaces, solid-immersion lenses, and cavities exhibit enhanced brightness and in some cases reduced linewidths, benefiting applications like quantum light sources~\cite{Hausmann2012-NanoLet, Hausmann2013-NanoLet}. Plasmonic coupling, particularly to metal nanoparticles or nanocolumns, has demonstrated directional intensity enhancement~\cite{hapuarachchi2024plasmonically}. These classical and quantum optics advances collectively show that NV emission is highly tunable through environmental engineering. However, most previous work isolated one factor, such as size, position, or environment, without a unified framework, prompting the need for integrative approaches that combine classical electrodynamics with quantum emitter dynamics. 
%\SQ{However, most previous work isolated one factor, such as size, position, or environment. Hybrid quantum–classical approaches have been used in other contexts for decades, but they have rarely been applied systematically to NV nanodiamonds. Here we present such a targeted framework, tailored to NV centres, that captures wavelength-dependent escape efficiency and the interplay of NV depth, nanodiamond size, and dielectric environment in a unified manner.}

Despite many theoretical studies on NV emission in nanodiamond, a predictive framework linking crystal size, NV position, and environment to radiative and non-radiative decay remains missing. Existing models often assume unit quantum efficiency or treat the NV as a two-level dipole, which fails in real nanodiamond systems where surface effects and dielectric mismatch play dominant roles~\cite{Chew1988-PRA}. A more complete treatment must account for the NV’s internal quantum dynamics alongside its classical dielectric environment. Practical design of NV devices is hindered by a lack of guidelines on what nanodiamond size or coating yields maximal emission without sacrificing spin coherence. 

Here we show a combined classical-quantum model that links NV electronic structure and nanodiamond geometry to emission brightness and escape efficiency under realistic conditions. Even though we did not incorporate the surface effects into our model, since the specifics of those effects can vary significantly from case to case, we focus on the computationally tractable general case of radiative decay, allowing us to isolate and quantify how the electromagnetic environment, specifically the dielectric structure surrounding the NV centre, modifies its emission properties. Our work employs a hybrid theoretical and computational approach that marries classical Maxwell’s equations for light scattering and dipole emission in and around the nanodiamond modelling with an open quantum systems model of the NV centre’s electronic-vibronic transitions. 
% Our work qualitatively and quantitatively reveals that the dielectric environment strongly modulates NV radiative rates and that higher surrounding index directly translates to brighter NV emission and shorter lifetimes. %Furthermore, nanodiamond size and NV position must be considered together, such as increasing ND size reduces electromagnetic losses but can introduce more non-radiative loss unless the NV is sufficiently far from surfaces. 

We demonstrate that nanodiamond size and NV position must be considered together: While increasing nanodiamond size supports emission enhancement, significant suppression can still occur unless the NV centre is sufficiently far from the surfaces. 
%This indicates that there should be an optimal regime where the ND is large enough to preserve NV quantum efficiency, yet not so large that it traps most photons. 
The findings in this work highlight %mark the note 
that a full electromagnetic and quantum analysis is necessary to predict subtle effects like wavelength-dependent outcoupling and the exact magnitudes of enhancement or suppression under various conditions.  While the examples presented here correspond to subwavelength nanodiamonds, the theoretical framework is a full-wave formulation that remains valid from the quasi-static to the wavelength-scale regimes. It captures both the near-field and radiative interactions without requiring the concept of propagating internal modes in the small-particle limit.

The remainder of this article is organised as follows. The next section details our theoretical framework, including the classical electromagnetic calculations of fields, the NV centre’s quantum-optical model%with rate equations
, and their combination to compute the emission spectra. Section 3 presents simulation results, analysing trends with nanodiamond size, NV position, and surrounding medium. Section 4 concludes with key findings and an outlook on modelling surface non-radiative effects and integrating NV nanodiamonds into hybrid photonic platforms. 

\section{Summary of the formalism}\label{Sec:Formalism}
\begin{figure*}[h!]
    \centering
    \includegraphics[width=0.75\textwidth]{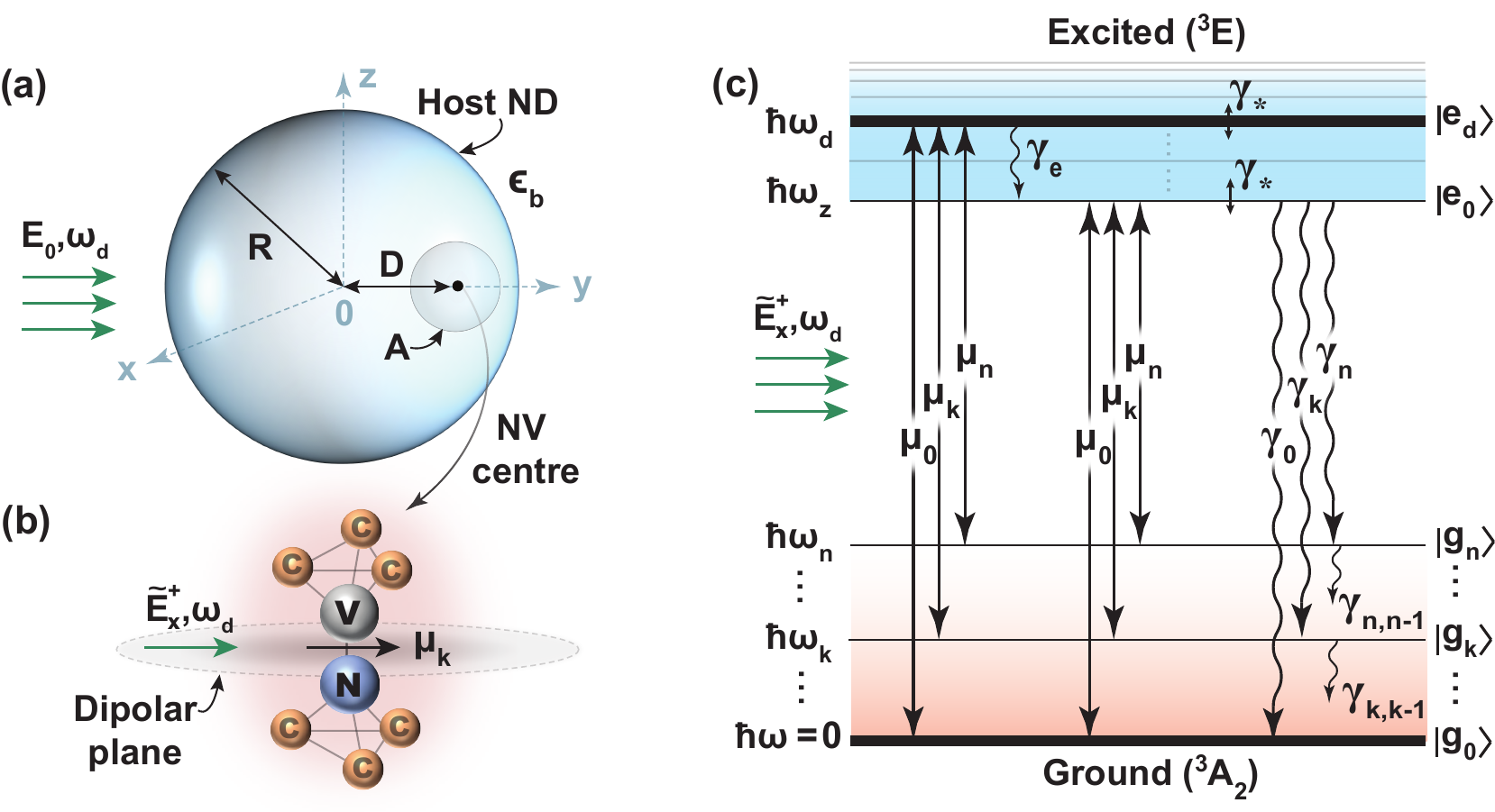}
    \caption{(Colour online) (a) Host nanodiamond (ND) in the medium of relative permittivity $\epsilon_b$ (b) Atomic structure and the orientation of the negatively charged nitrogen-vacancy (NV) centre embedded in the ND (c) Optical abstraction of the NV level structure.}
    \label{Fig:Schematic_and_LevelDiag}
\end{figure*}

Throughout this work, we consider a spherical ND of radius $R$ hosting a single negatively-charged NV at a distance $D$ from the origin along the $x-$axis, as depicted in Fig.\ \ref{Fig:Schematic_and_LevelDiag}(a). The N$-$V axis is oriented parallel to the $z-$axis of the Cartesian coordinate system considered, and the plane of the optical dipole formation of the NV centre coincides with the $xy$ plane. The ND is surrounded by a non-absorptive dielectric background medium of relative permittivity $\epsilon_b$. A driving field with an angular frequency $\omega_d$ exceeding that of the NV zero phonon line (such as $\SI{532}{\nano\meter}$ green light), polarised along the $x$ axis, is incident on the ND, through the background medium. This field takes the form $E=E_0e^{-i\omega_d t}+c.c.$ where $t$ denotes time, and $c.c.$ denotes the complex conjugate of the preceding expression, respectively.

% \st{In practice, NDs are rarely truly spherical}~\cite{Reineck2019-PPSC, Eldemrdash2023-Carbon}\st{. Despite the complexities of modeling misshapen NDs, a spherical approximation captures many of the key optical effects present in them.}%Although such misshapen NDs are difficult to model, many of the same optical effects will be present, as are considered in our spherical model.
In practice, nanodiamonds are rarely spherical and often display disk- or rod-like geometries~\cite{Reineck2019-PPSC, Eldemrdash2023-Carbon}. We adopt a spherical approximation to provide a tractable baseline that isolates the fundamental electrodynamic effects of size, NV position, and dielectric environment. Although this approximation cannot reproduce shape-specific resonances, it captures the leading-order changes in local density of states and emission efficiency. Deviations in real geometries may change absolute intensities by factors of order unity, but the overall trends identified here remain robust. The simulation framework described can be extended to more complex nanodiamond shapes by replacing Mie theory with numerical solvers such as finite-difference time-domain~\cite{KaneYee1966} or surface integral methods~\cite{Sun2020, Sun2022} for solving Maxwell’s equations.

%\begin{figure*}[h!]
%    \centering
%    \includegraphics[width=0.75\textwidth]{Figures/Level_Diag_v3.pdf}
%    \caption{(Colour online) (a) Host nanodiamond (ND) in the medium of relative permittivity $\epsilon_b$ (b) Atomic structure and the orientation of the negatively charged nitrogen-vacancy (NV) centre embedded in the ND (c) Optical abstraction of the NV level structure.}
%    \label{Fig:Schematic_and_LevelDiag}
%\end{figure*}

\subsection{Green input field at the NV location}\label{Sec:Field_mod}
The NV center embedded inside does not experience the simple form of the electric field incident on the ND. Instead, it experiences a localised field that is a complex superposition resulting from the interplay between several phenomena related to the interaction of the incident field with the ND and the surrounding dielectric medium. These phenomena are briefly described below.
 
The incident field undergoes scattering and transmission at the interface between the background medium and the ND. According to Maxwell's equations, only a fraction of the incident field is transmitted into the ND, with a modified amplitude dependent on the refractive indices of the two media~\cite{bohren2008absorption}. Furthermore, the high refractive index of diamond not only reduces the speed and wavelength of light within the nanodiamond, but also leads to strong local field screening effects that modify the electric field experienced by the NV center.

Collectively, these effects are rigorously captured by Maxwell’s equations, supplemented with continuous conditions of the tangential components of electric and magnetic fields across boundaries. In this work, we set the exciting green light incident on the NV as a linearly polarized plane wave, in which the green light field at the NV location are solved by Mie theory~\cite{Mie1908-AdP, Van_de_Hulst1981-hj}. %Solving these equations yields a resultant electromagnetic field projection along the dipolar plane (see Fig.~\ref{Fig:Schematic_and_LevelDiag}), at the NV location, which we denote as $E_\text{NV}=\tilde{E}^+_xe^{-i\omega_g t}+c.c.$, where $\tilde{E}^+_x$ is the modified complex amplitude. 
After solving these equations, we extract the projection of the electromagnetic field along the NV dipolar plane (see Fig.~\ref{Fig:Schematic_and_LevelDiag}), at the NV location, which we denote as $E_\text{NV}=\tilde{E}^+_xe^{-i\omega_d t}+c.c.$, where $\tilde{E}^+_x$ is the modified complex amplitude.% and $\omega_d$ is the angular frequency of the driving field. 
%\TODO{SQ to revise as needed and add the details of simulation procedure}. \SQ{done}

\subsection{Decay rate modification}\label{Sec:Decay_mod}
In addition to modifying the complex electric field amplitude of the green input field at the NV location, the surrounding electromagnetic environment also influences the spontaneous decay rates of the NV emission bands. This occurs through changes to LDOS \cite{hohenester2019nano}, which directly impact the radiative decay channels available to the NV.

We treat the decay rate $\gamma_k$ of the $k$th NV emission band as,
\begin{equation}\label{Eq:Decay_mod}
    \gamma_k\approx\varepsilon_k\cdot\gamma_\text{ref}\cdot\frac{\gamma_u(\lambda_k)}{\gamma^\text{ref}_u(\lambda_k)},
\end{equation}
where $\gamma_\text{ref}$ (approximated as $1/\SI{12}{\nano\second}$~\cite{Doherty2013-PR}) is the total decay rate of an NV centre in the reference environment taken to be bulk diamond and $\varepsilon_k$ is the approximate fraction of photons decaying through the $k$th NV emission band such that $\sum_k\varepsilon_k=1$, as detailed in section~\ref{Sec:Quantum_Model}. Therefore, $\varepsilon_k\gamma_\text{ref}$ represents the estimated decay rate of the $k$th NV emission band in the reference environment (bulk diamond).

The ratio $\gamma_u(\lambda_k)\big/\gamma^\text{ref}_u(\lambda_k)$ is the decay rate modification of a unit dipole placed at the NV location relative to a unit dipole in the reference environment. This is computed as,
\begin{equation}\label{Eq:Decay_mod_from_power}
    \frac{\gamma_u(\lambda_k)}{\gamma^\text{ref}_u(\lambda_k)}=\frac{P(\lambda_k)/(\hbar\omega_k)}{P_\text{ref}(\lambda_k)/(\hbar\omega_k)}
\end{equation}
where $P(\lambda_k)$ is the rate of energy dissipation by a unit dipole of wavelength $\lambda_k$ computed by integrating the Poynting vector over a small closed surface area A with a radius of 1 nm (marked in Fig.\ \ref{Fig:Schematic_and_LevelDiag}) enclosing the NV centre within the nanodiamond \cite{issa2007fluorescence, Dolde2011-NatPhys, Teissier2014-PRL, Poggiali2017-PRB}. $P(\lambda_k)/(\hbar\omega_k)$, where $\omega_k=2\pi c/\lambda_k$ with $c$ being the speed of light, is therefore the estimated rate of photon emission from the unit dipole. $P_\text{ref}(\lambda_k)/(\hbar\omega_k)$ similarly estimates the rate of photon emission from a unit dipole in the reference environment. The Poynting vector is obtained from the electromagnetic fields via solving the Maxwell's equations using Mie theory~\cite{Mie1908-AdP, Margetis2002-JMP, Sun2023-PRA} due to the radiation of a unit dipole at the NV location.

%\TODO{SQ to revise as needed and add the details of simulation procedure}.  \SQ{done}

% In practical systems, additional nonradiative decay channels may be present, often arising from sources such as surface adsorbates, chemical terminations, or structural defects. These channels could potentially act as absorbers or perturb the local charge environment around the NV centre, leading to emission quenching. Since the specifics of these effects can vary significantly from case to case, they are not incorporated into our model. Instead, we focus on the computationally tractable general case of radiative decay, allowing us to isolate and quantify how the electromagnetic environment, specifically the dielectric structure surrounding the NV centre, modifies its emission properties.

\subsection{Quantum optical model and the near-field spectrum}\label{Sec:Quantum_Model}

We use the modified green electric field $E_\text{NV}$ and decay rates $\gamma_k$ computed as above to estimate the near-field emission spectrum of the NV centre, radiated within the nanodiamond. For this, we adopt a validated room-temperature quantum optical abstraction of the NV centre developed in \cite{hapuarachchi2022nv, SGH08}, a summary of which is provided below. The corresponding level diagram is shown in Fig.~\ref{Fig:Schematic_and_LevelDiag}(c). Our model employs room-temperature parameters extracted experimentally~\cite{albrecht2013coupling}, which phenomenologically captures the broadening and redistribution of emission intensity across phonon sidebands. While this approach accounts for the dominant impact of phonon interactions at ambient conditions, it does not explicitly model the full thermal population of vibrational states. Such a treatment would further refine the spectral distribution, but lies outside the scope of this baseline framework.

In this abstraction, the NV center is modelled as a multi-level system comprising $n+1$ optical ground states $\lbrace\ket{g_k}\rbrace$, with energies $\lbrace\hbar\omega_k\rbrace$, where $k\in\lbrace0,\hdots,n\rbrace$ with $\hbar \omega_0 = 0$, and two effective optical excited states $|e_0\rangle$ and $|e_d\rangle$ with energies $\hbar\omega_z$ (the zero-phonon line energy), and $\hbar\omega_d$, respectively. The subscript $k$ indicates the number of phonon quanta separating $\ket{g_k}$ from the lowest ground level $\ket{g_0}$. The lowest-energy optical excited state is denoted by $|e_0\rangle$. The higher-energy phononic excitations above $|e_0\rangle$ are collectively represented by a phenomenologically defined upper excited level $|e_d\rangle$, which resonantly couples to the driving field at the NV location.

We assume that the NV center undergoes coherent transitions $|e_j\rangle \leftrightarrow |g_k\rangle$ ($j\in\lbrace 0,d\rbrace$ and $k\in\lbrace0,...,n\rbrace$) upon the incidence of the driving field ($E_\text{NV}=\tilde{E}^+_xe^{-i\omega_d t}+c.c.$) polarised along the $x-$axis, with transition dipole elements $\mu_k$ ($k\in\lbrace 0,n\rbrace$) for each optical transition aligned along incident polarisation.  An ultrafast nonradiative (phononic) decay $\ket{e_d}\to\ket{e_0}$ occurs at a rate $\gamma_\text{e}$. This combination of the coherent excitation followed by rapid phononic decay to the lower excited state $|e_0\rangle$ effectively acts as an incoherent pumping mechanism for the NV centre. At room temperature, strong electron-phonon interactions suppress coherent oscillations, and the dynamics of population transfer are well described by an effectively incoherent process. We use the Lindblad formalism to maintain generality and consistency across different regimes; however, in the parameter range studied here, the results are equivalent to those obtained from a simpler rate-equation model.

We further assume that, for the nanodiamond sizes considered, the emission dipoles remain aligned along the $x-$axis, preserving the directionality of emission \cite{christinck2020characterization}. A pure dephasing rate $\gamma_*$ accounts for random fluctuations of the excited states relative to the ground levels. The incoherent optical transition $|e_0\rangle\to|g_0\rangle$ contributes to the \emph{zero phonon line (ZPL)} of the emission spectrum, while transitions $|e_0\rangle\to|g_k\rangle$ (for $k \neq 0$) contribute to the phonon sidebands. Nonradiative transitions between adjacent ground states $|g_k\rangle\to|g_{k-1}\rangle$ occur at rates $\gamma_{k,k-1}$ for $k\in\lbrace 1,\ldots,n\rbrace$. 

The Hamiltonian of the NV center, expressed in a reference frame rotating at the angular frequency of the driving field ($\omega_d$), takes the following form \cite{hapuarachchi2022nv},
\begin{equation}\label{Eq:RF_Hamiltonian}
	\hat{H}_\text{\tiny{RF}} \approx \left(\sum_{k=0}^n \hbar\omega_k|g_k\rangle\langle g_k|\right) + \hbar(\omega_\text{z} - \omega_d)|e_0\rangle\langle e_0| - \sum_{j\in\lbrace0,d\rbrace}\sum_{k=0}^n\left(\hbar\Omega_k^\text{r}|e_j\rangle\langle g_k| + \hbar\Omega_k^\text{r*}|g_k\rangle\langle e_j| \right),
\end{equation}
where the derivation employs the rotating wave approximation (RWA), in which fast-oscillating terms that average to zero over the population oscillation timescales of interest are neglected. In this expression, $\Omega_k^\text{r}$ is the complex Rabi frequency that drives the $\ket{e_j} \leftrightarrow \ket{g_k}$ transition, and $\Omega_k^{\text{r}*}$ is its complex conjugate. Each Rabi frequency is related to the complex amplitude $\tilde{E}^+_x$ of the green electric field at the NV location via $\hbar \Omega_k^\text{r} = \mu_k \tilde{E}^+_x$.

We then estimate the evolution of the quantum state of the NV centre (represented by the density operator $\rho$ in the same reference frame as the Hamiltonian) as an open quantum system using the following form of the Lindblad master equation \cite{breuer2002theory, campaioli2024quantum, hapuarachchi2022nv, hapuarachchi2024plasmonically},
\begin{equation}\label{Eq:Density_matrix_master_eq}
\dot{\hat{\rho}} = -\frac{i}{\hbar}[\hat{H}_\text{\tiny{RF}}, \hat{\rho}] + \sum_{l}\Gamma_l\left(\hat{L}_l^{\phantom{\dagger}}\hat{\rho}\hat{L}_l^\dagger -\frac{1}{2}\lbrace \hat{L}_l^\dagger \hat{L}_l^{\phantom{\dagger}}, \hat{\rho}\rbrace\right),
\end{equation}
where $\hat{L}_l$ is the Lindblad or collapse operator corresponding to the $l$th decoherence channel with characteristic decoherence rate $\Gamma_l$. The mathematical operators $[\cdot,\cdot]$ and $\lbrace\cdot,\cdot\rbrace$ denote the commutator and anti-commutator of the operands. The decoherence channels and the corresponding rates considered in our model are summarised in Table \ref{Table:Operators}.

\begin{table}[!ht]
	\begin{center}
		\begin{tabular}{ |l|l|l| } 
			\hline
			Transition(s) & Rate $\Gamma_l$ & Operator $\hat{L}_l$ \\
			\hline
			Optical decay $|e_0\rangle \to |g_k\rangle$, for $k\in\lbrace 0,\hdots,n\rbrace$  & $\gamma_k$     & $\hat{\sigma}_k = |g_k\rangle\langle e_0|$  \\ 
			Phononic decay $|g_k\rangle \to |g_{k-1}\rangle$, for $k\in\lbrace 1,\hdots,n\rbrace$ & $\gamma_{k,k-1}$  & $|g_{k-1}\rangle\langle g_k|$ \\ 
			Ultrafast nonradiative decay $|e_d\rangle \to |e_0\rangle$ & $\gamma_\text{e}$  & $|e_0\rangle\langle e_d|$ \\
			Dephasing of the excited states relative to the ground & $\gamma_*$   & $|e_0\rangle\langle e_0|+|e_d\rangle\langle e_d|$ \\ 
			\hline
		\end{tabular}
		\caption{Operators and decoherence rates for the Lindblad equation (\ref{Eq:Density_matrix_master_eq})}
		\label{Table:Operators}
	\end{center}
\end{table}

%\begin{flushleft}
%	For each optical decay transition $|e_0\rangle \to |g_k\rangle$:\\
%	$\Gamma_l = \gamma_k$, for $k\in\lbrace 0,\hdots,n\rbrace$\\
%	$\hat{L}_l = \hat{\sigma}_k = |g_k\rangle\langle e_0|$
%\end{flushleft}
%
%\begin{flushleft}
%	For each phononic decay transition $|g_k\rangle \to |g_{k-1}\rangle$:\\
%	$\Gamma_l = \gamma_{k,k-1}$, for $k\in\lbrace 1,\hdots,n\rbrace$\\
%	$\hat{L}_l = |g_{k-1}\rangle\langle g_k|$
%\end{flushleft}
%
%\begin{flushleft}
%	Ultrafast nonradiative decay in the excited state $|e_1\rangle \to |e_0\rangle$:\\
%	$\Gamma_l = \gamma_\text{e}$\\
%	$\hat{L}_l = |e_0\rangle\langle e_1|$
%\end{flushleft}

%\begin{flushleft}
%	Dephasing from excited to ground states:\\
%	$\Gamma_l = \gamma_*$\\
%	$\hat{L}_l = |e_0\rangle\langle e_0|+|e_1\rangle\langle e_1|$
%\end{flushleft}

The steady state density matrix obtained by solving the master equation above can be used to estimate the total photon emission intensity spectrum in the near-field as follows \cite{campaioli2024quantum, hapuarachchi2022nv, hapuarachchi2024plasmonically},
\begin{equation}\label{Eq:Total_emission_intensity}
	I_\text{nf}(\omega) \propto \sum_{k=0}^{n} \gamma_k \int_{-\infty}^{\infty}d\tau e^{-i\omega\tau}\langle\hat{\sigma}_k^\dagger(\tau)\hat{\sigma}_k(0)\rangle_\text{ss},
\end{equation}
where $\langle\cdot\rangle_\text{ss}$ denotes the expectation of the operator within, computed using the steady state density matrix. $I_\text{nf}(\omega)$ is designated as the near-field spectrum radiated within the nanodiamond, as we used a surface area (A) within the ND for the Poynting vector integral in section~\ref{Sec:Decay_mod}.
\\\\
\textbf{Parameters of the quantum optical model}: Throughout this work, we use the experimentally informed spectral parameters extracted by Albrecht \emph{et al.} in \cite{albrecht2013coupling} when quantum optically modelling an NV center at room temperature. These parameters have been obtained by fitting an experimentally measured NV emission intensity spectrum with 8 Lorentzian lines ($n=7$) with scaled amplitudes. These amplitudes $A_k$ where $\varepsilon_k = A_k\big/\sum_k A_k$, phonon energies $\hbar\omega_k$, and phonon decay rates $\gamma_{k,k-1}$ are presented in Table\ \ref{Table:Params_from_Roland}. 

\begin{table}[!ht]
	\begin{center}
		\begin{tabular}{ |c|c|c|c| } 
			\hline
			k & $A_k$ (arb.u) & $\hbar\omega_k$ $(\SI{}{\milli\electronvolt})$ & $\gamma_{k,k-1}$ $(\SI{}{\tera\hertz}$) \\
			\hline
			0 & 1520  & 0     & -  \\ 
			1 & 5260  & 31.8  & 85 \\ 
			2 & 18600 & 70.3  & 82 \\
			3 & 16400 & 124   & 79 \\ 
			4 & 14000 & 168   & 88 \\ 
			5 & 9180  & 221   & 65 \\
			6 & 6570  & 275   & 71 \\ 
			7 & 3270  & 319   & 86 \\
			\hline
		\end{tabular}
		\caption{Room temperature NV parameters from \cite{albrecht2013coupling}.}
		\label{Table:Params_from_Roland}
	\end{center}
\end{table}

The energy of the NV zero-phonon line $\hbar\omega_\text{z} = \SI{1.941}{\electronvolt}$ \cite{albrecht2013coupling}, dephasing rate between the ground and excited states $\gamma_*=\SI{15}{\tera\hertz}$ \cite{albrecht2013coupling}, phononic decay rate between the two excited states $\gamma_e\sim1434$ THz \cite{hapuarachchi2024plasmonically}, $\mu_0\sim\SI{5.2}{D}$ (the dipole moment element that corresponds to the $|e_j\rangle\leftrightarrow|g_0\rangle$ transition for both $j=0$ and $1$) \cite{alkauskas2014first} and $\mu_k=\mu_0 \sqrt{\varepsilon_k/\varepsilon_0}$ \cite{hapuarachchi2022nv} is used throughout our study.

\subsection{Far field emission spectrum}\label{Sec:Far_field_spectra}

Having characterized the near-field emission spectrum within the nanodiamond, the subsequent step involves determining the far-field emission spectrum. This refers to the spectrum observable by a detector positioned remotely within the infinitely extending background medium. The process by which light generated internally by the NV center propagates from the nanodiamond into the far field is profoundly influenced by the NV location and polarization, the optical properties and size of the nanodiamond itself, as well as the surrounding dielectric environment. This transition, similar to the modification of the externally incident light when reaching the NV center, involves a complex interplay of several optical phenomena.

The finite size and spherical geometry of the nanodiamond lead to internal scattering that redistributes the emitted power, transforming the intrinsic dipole-like emission pattern of the NV center into a complex, wavelength-dependent far-field distribution that varies with the ND size and NV location within it. As such, the background dielectric medium plays a direct role in shaping these effects and, ultimately, the far-field spectrum. 

To simulate this process, the NV center is represented by an effective electric dipole the strength and orientation of which are determined by the local optical environment. The dipole strength, $|\mathbf{p}|$, is related to the near-field emission profile of the NV center, $I_\text{nf}(\lambda)$, calculated in Eq.~(\ref{Eq:Total_emission_intensity}) as~\cite{Robledo2010-PRL, Novotny2012}:
\begin{align}
    |\mathbf{p}| \sim \sqrt{I_\text{nf}(\lambda) \lambda^3}.
\end{align}
Here, $I_\text{nf}(\lambda)$ incorporates the full dependence on the excitation-field polarization, NV location, and nanodiamond size described earlier. In our simulations for the far-field emission profiles, the emission dipole $\mathbf{p}$ is aligned with the polarization of the excitation light. The far-field radiation is then obtained by solving Maxwell’s equations explicitly, using $\mathbf{p}$ as the source dipole~\cite{Mie1908-AdP, Margetis2002-JMP, Sun2023-PRA}, which fully accounts for the NV orientation, position, and nanodiamond size.

%To simulate this process, the NV center is represented by an electric dipole. The dipole strength, $|\mathbf{p}|$, and the modified near field emission profile of NV calculated in Section~\ref{Sec:Quantum_Model} has the following relationship~\cite{Robledo2010-PRL, Novotny2012}:
%\begin{align}
%    |\mathbf{p}| \sim \sqrt{\Gamma \lambda^3}
%\end{align}
%in which $\Gamma$ represents the near field emission profile of NV. Also, the direction of $\mathbf{p}$ is set the same as the Green light polarisation. Similar to the procedure in Section~\ref{Sec:Decay_mod}, the far-field radiation of NV is simulated by solving the Maxwell's equations with the Mie theory~\cite{Mie1908-AdP, Margetis2002-JMP, Sun2023-PRA}.

%\TODO{SQ to revise as needed and add the details of simulation procedure}. \SQ{done}

\section{Results and discussion}\label{Sec:Results_and_discussion}

\subsection{Green input field at the NV location}\label{Sec:Field_mod_Results}

\begin{figure*}[!ht]
    \centering
    \includegraphics[width=0.8\textwidth]{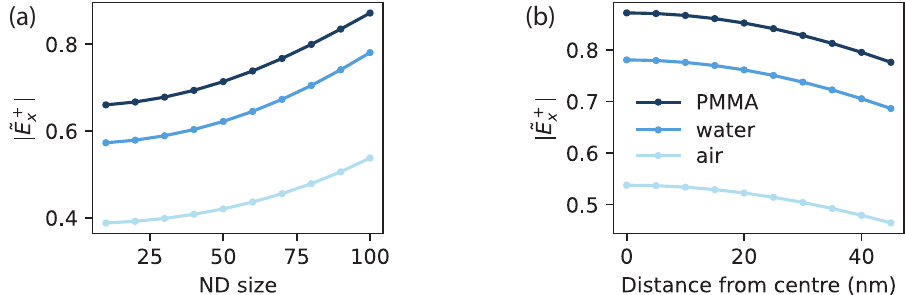}
    \caption{(Colour online) Fraction of $x$-polarised electric field amplitude at the NV location relative to the incident green light, $|\tilde{E}_x^+|$. (a) $|\tilde{E}_x^+|$ as a function of nanodiamond diameter ($2R$), with the NV center fixed at the origin ($D=0$), showing the effect of a different infinitely extending background dielectric media (air with refractive index of $n=1$, water $n=1.33$, PMMA $n=1.5$). (b) $|\tilde{E}_x^+|$ as a function of the radial distance to the NV location from the origin along the $x$-axis, for a $\SI{100}{\nano\meter}$ diameter nanodiamond, for the same background media.}
    \label{Fig:Classical_E_field_mod}
\end{figure*}

The numerically computed fraction of $x$-polarised electric field amplitude at the NV location relative to the incident green light, $|\tilde{E}_x^+|$ is presented in Fig.~\ref{Fig:Classical_E_field_mod}. These results display the significant modification of the incident field due to the surrounding dielectric environment, the ND geometry, and NV location, consistent with the phenomena discussed in section~\ref{Sec:Field_mod}.

Fig.~\ref{Fig:Classical_E_field_mod}(a) shows the variation of $|\tilde{E}_x^+|$ against ND size ($2R$), with the NV center fixed at the origin of the spherical ND ($D=0$), for different infinitely extending background dielectric media air (refractive index $n_b=1$), water ($n_b=1.33$), and PMMA ($n_b=1.5$). Fig.~\ref{Fig:Classical_E_field_mod}(b) shows the variation of $|\tilde{E}_x^+|$ as the NV center is moved along the $x-$axis from the origin of a $\SI{100}{\nano\meter}$ diameter nanodiamond ($n_D=2.4$), for the same three background media. 

For all ND sizes, the field amplitude experienced by the NV is highest when the ND is embedded in PMMA, followed by water, and then air. This trend is directly related to the refractive index contrast between the background medium and the nanodiamond which has a high refractive index $n_D=2.4$. As the refractive index of the background $n_b$ increases and approaches that of diamond, this results in reduced dielectric screening, increasing the field experienced at the NV location with $n_b$.

For a given background medium, $|\tilde{E}_x^+|$ tends to exhibit a non-linear dependence on the ND size, which is likely to arise due to scattering effects. As the ND size increases, its internal field distribution tends to increase in complexity due to interference. While larger NDs can capture more incident light, these internal interference effects can lead to varying field strengths dependent on the exact NV-location within the sphere. For all background media, $|\tilde{E}_x^+|$ decreases as the NV location moves away from the origin of the nanodiamond towards the surface, which is likely to be a consequence of the internal field distribution governed by scattering and screening effects. Such an excitation reduction lowers the NV emission intensity. 

%In addition to these surface-induced excitation reductions, band bending effects near the nanodiamond surface can also reduce the NV center’s expected average brightness. This band bending can shift the NV charge-state equilibrium or alter the internal field environment, affecting excitation efficiency or emission rates~\cite{Hauf2011-PRB, Hu2013-CTC}. While our model does not account for band bending, the emission suppression we observe due to radiative and non-radiative pathways is expected to occur alongside these electronic effects in real systems.

%\TODO{SQ to revise section as preferred}. \SQ{Reads good}

\subsection{Decay rate modification}\label{Sec:Decay_mod_results}

\begin{figure*}[!ht]
    \centering
    \includegraphics[width=\textwidth]{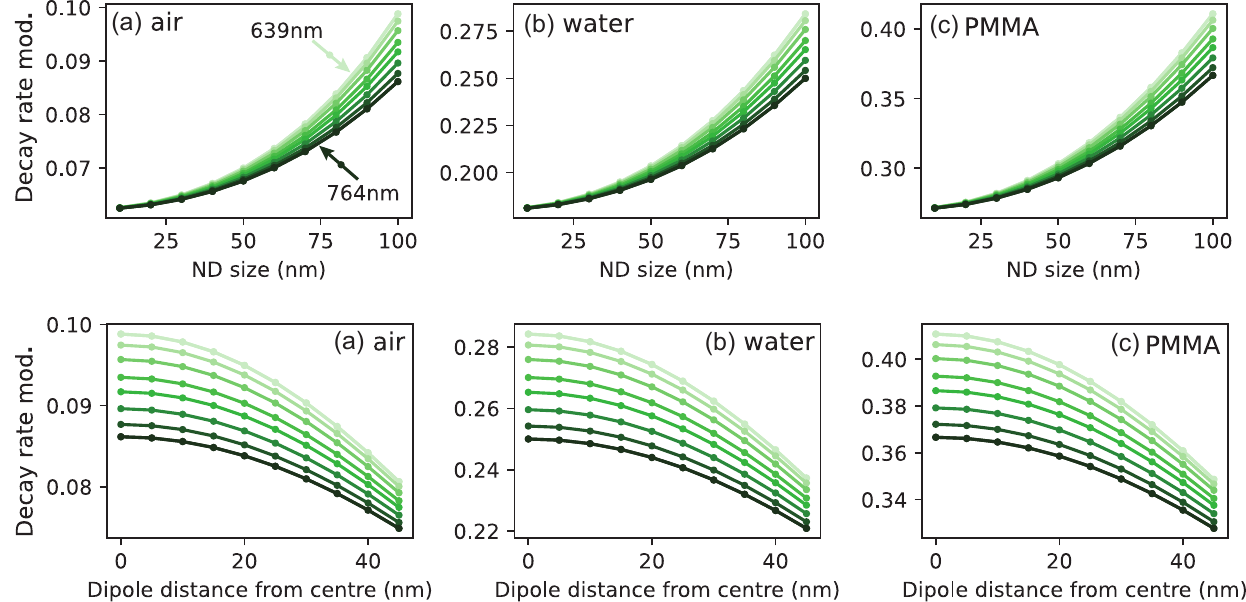}
    \caption{(Colour online) Numerically computed decay rate modifications, $\gamma_u(\lambda_k)/\gamma^\text{ref}_u(\lambda_k)$, for unit dipoles emitting at NV transition wavelengths $\lambda_k\in\lbrace 639, 649, 663, 682, 699, 721, 744, 764 \rbrace\;\SI{}{\nano\meter}$ within a spherical nanodiamond, relative to a unit dipole in infinitely extending bulk diamond (reference). The first row depicts $\gamma_u(\lambda_k)/\gamma^\text{ref}_u(\lambda_k)$ as a function of nanodiamond size ($2R$) for a unit dipole located at the origin of the spherical nanodiamond ($D=0$) when the infinitely extending background medium is (a) air, (b) water and (c) PMMA. The second row depicts decay rate modification as a function of the unit dipole location along the $x-$axis in a $\SI{100}{\nano\meter}$ diameter nanodiamond in background media (d) air (e) water and (f) PMMA. The legend in (a) is common for all subplots and colors of the lines progressively darken with increasing emission wavelength (from $\SI{639}{\nano\meter}$ to $\SI{764}{\nano\meter}$).}
    \label{Fig:Classical_decay_mod}
\end{figure*}

Fig.~\ref{Fig:Classical_decay_mod} shows decay rate modifications $\gamma_u(\lambda_k)/\gamma^\text{ref}_u(\lambda_k)$ for unit dipoles inside a spherical nanodiamond relative to bulk diamond as the reference environment, computed using Poynting vector integrals inside the diamond, as discussed in section~\ref{Sec:Decay_mod}. These values mimic the local density of optical states (LDOS) \cite{hohenester2019nano} experienced by a unit dipole at the NV location within the nanodiamond, relative to bulk diamond. The modification factor remains $<1$ for all cases, indicating that the LDOS experienced by a dipole within the finite nanodiamond is dramatically suppressed compared to a dipole in infinite bulk environment. Our simulations indicate that this suppression is more pronounced for longer emission wavelengths (darker lines).

The first row of Fig.~\ref{Fig:Classical_decay_mod} (subfigures a-c) shows the LDOS dependence on nanodiamond size for a dipole at the origin of the spherical nanodiamond submerged in infinitely extending air, water, and PMMA backgrounds. As the background refractive index increases (from air to PMMA), the overall LDOS suppression significantly decreases as the reduced refractive index contrast between diamond and the background creates a more bulk-like internal electromagnetic environment. 

The second row of Fig.~\ref{Fig:Classical_decay_mod} (subfigures d-f)  depicts the LDOS dependence on dipole location within a $\SI{100}{\nano\meter}$ diameter nanodiamond in the same background media. For all submerging media, LDOS decreases as the dipole approaches the surface, highlighting the strong influence of the dielectric interface on emission dynamics. 

\subsection{Near-field spectra}\label{Sec:Near_field_results}

\begin{figure*}[!ht]
    \centering
    \includegraphics[width=\textwidth]{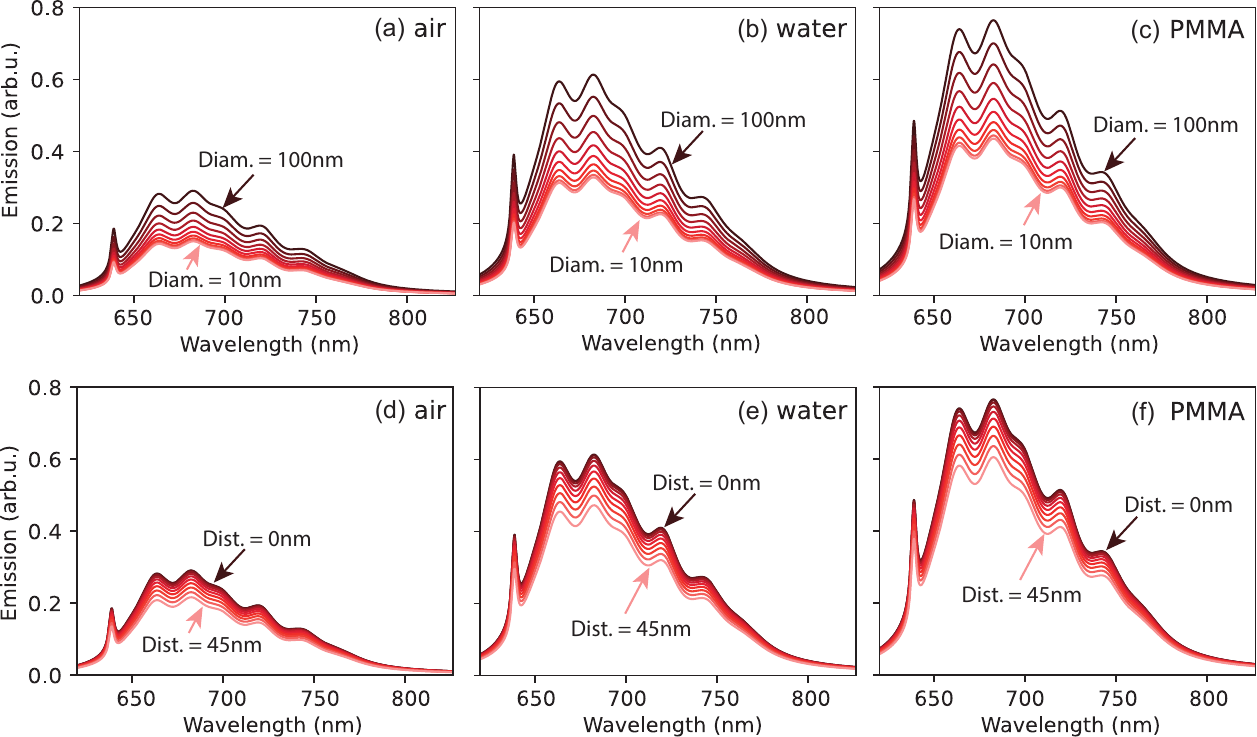}
    \caption{(Colour online) Numerically computed near-field emission spectra for an NV center within a spherical nanodiamond, radiated inside the nanodiamond, as described in Section~\ref{Sec:Quantum_Model}. The first row depicts the variation of emission intensity spectra of an NV center located at the origin ($D=0$) as a function of nanodiamond diameter ($2R$), ranging from $\SI{10}{\nano\meter}$ to $\SI{100}{\nano\meter}$ (line colors darken with \emph{increasing} diameter). Each column corresponds to a different infinitely extending background medium, (a) air, (b) water, and (c) PMMA. The second row depicts the variation of emission intensity spectra for different NV distances from the origin along the $x-$axis (ranging from $\SI{0}{\nano\meter}$ to $\SI{45}{\nano\meter}$, with line colors darkening with \emph{decreasing} distance) in a $\SI{100}{\nano\meter}$ diameter nanodiamond. All plots are normalised by the maximum intensity of the emission spectrum of an NV centre in an infinitely extending bulk diamond as the reference.}
    \label{fig:Near_field}
\end{figure*}

Fig.~\ref{fig:Near_field} presents the numerically computed near-field emission spectra for an NV center in a spherical nanodiamond, for different nanodiamond sizes, NV locations, and background media, radiated within the nanodiamond as discussed in section~\ref{Sec:Quantum_Model}. These spectra are normalised by the maximum intensity of the emission spectrum of an NV centre in an infinitely extending bulk diamond (reference).

Fig.~\ref{fig:Near_field}(a)-(c) in the first row depict the near-field spectra as a function of nanodiamond diameter for NV centres at the origin of the spherical nanodiamond hosts, immersed in air, water, and PMMA, respectively. A general trend observed across all background media is that increasing nanodiamond diameter leads to a substantial increase in the overall near-field emission intensity (with darker lines corresponding to larger diameters showing higher intensities). This behavior is largely contributed by the increasing strength of the optical electric field at the NV location as the nanodiamond size increases, as detailed in Fig.~\ref{Fig:Classical_E_field_mod}(a). While decay rate modifications also vary with size (Fig.~\ref{Fig:Classical_decay_mod}(a-c)), our analyses suggest that their relative impact on the overall intensity variation is secondary compared to the impact of the excitation field. For all sizes, the highest emission intensities are observed when the nanodiamond is immersed in PMMA, followed by water, and then air. This directly correlates with the reduced dielectric screening in the background media with refractive indices closer to that of diamond, leading to a stronger excitation field at the NV. Panels (d)-(f) in the second row of Fig.~\ref{Fig:Classical_decay_mod} show the near-field emission intensity spectra as the NV center is moved along the $x-$axis from the origin towards the surface of a $\SI{100}{\nano\meter}$ diameter nanodiamond in air, water, and PMMA, respectively. 
% \st{In all background media, the near-field emission intensity radiated within the nanodiamond host decreases as the NV center moves away from the origin towards the nanodiamond surface (as darker lines corresponding to smaller distances from the center show higher intensities)}. \SQ{In all background media, the near-field emission intensity decreases as the NV center moves away from the origin towards the surface. While the overall spectral profile remains similar, close inspection reveals wavelength-dependent differences in the phonon sideband structure. For example, the relative height of the 682 nm component compared to shorter-wavelength sidebands varies with ND size and NV position. We highlight these subtle variations in Fig. 4 by including insets/difference spectra.} \HH{Should we emphasise wavelength dependent differences in the far-field spectra (Fig. 5) instead?}\st{This reduction is contributed by the decrease in the magnitude of the green excitation field experienced by the NV center as it approaches the surface, as observable in Figure}~\ref{Fig:Classical_E_field_mod}(b). 

Careful observation of Fig.~\ref{fig:Near_field} reveals that the near field spectra exhibit different trends with varying ND diameter and NV position, in addition to changes in the overall intensity. For example, as the ND diameter increases, an NV center at the origin exhibits increasingly larger emission enhancements (where the gap between adjacent plots increase). However, as we can observe from the second row, as the NV centre moves towards the origin of a fixed sized nanodiamond, the gap between adjacent plots tends to decrease while the emission intensity increases.

\subsection{Far-field spectra}\label{Sec:Far_field_results}

\begin{figure*}[!ht]
    \centering
    \includegraphics[width=\textwidth]{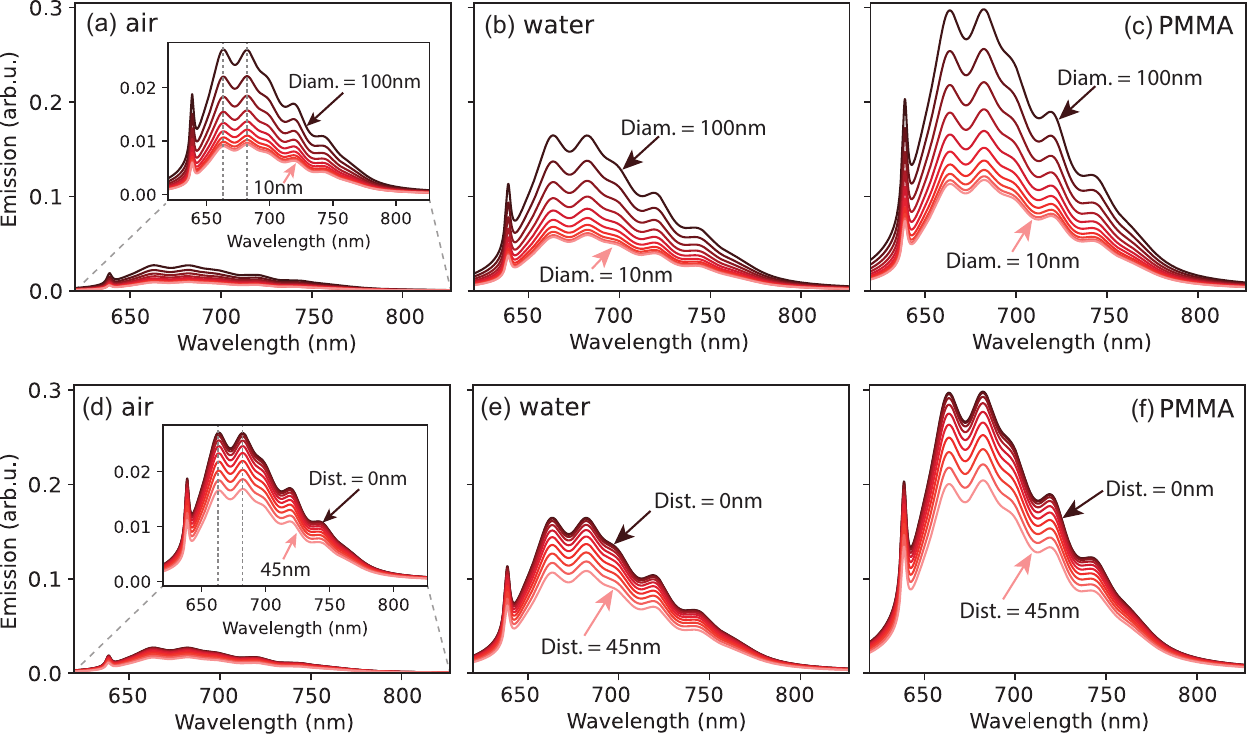}
    \caption{(Colour online) Numerically computed far-field emission intensity spectra for an NV center hosted within a spherical nanodiamond. The first row shows the variation of far-field spectra with nanodiamond diameter ($2R$, where darker lines correspond to increasing diameters), for an NV centre located at the origin of the nanodiamond host submerged in background media (a) air, (b) water, and (c) PMMA. The second row shows the variation of far-field spectra with the distance to the NV location from the origin along the $x-$axis (ranging from $\SI{0}{\nano\meter}$ to $\SI{45}{\nano\meter}$, with darker lines corresponding to decreasing distance) in a $\SI{100}{\nano\meter}$ diameter nanodiamond in (d) air, (e) water, and (f) PMMA. Insets in (a) and (d) provide an enlarged view of the y-axis to highlight details of the significantly suppressed emission in air. All plots are normalised by the maximum intensity of the emission spectrum of an NV centre in an infinitely extending bulk diamond as the reference.}
    \label{fig:Far_field}
\end{figure*}

Having analysed the variation of near-field emission spectra in section~\ref{Sec:Near_field_results}, we present the corresponding far-field emission spectra, which represent the light that has escaped the nanodiamond host and is ultimately observable by a detector positioned remotely within the infinitely extending background medium. Fig.~\ref{fig:Far_field} displays these far-field spectra, for the same parameter variations (nanodiamond size, NV location, and background media) and panel layout as the near-field plots. All far-field spectra are normalised by the maximum intensity of the emission spectrum of an NV centre in infinitely extending bulk diamond (reference).

A direct comparison of the far-field spectra in Fig.~\ref{fig:Far_field} with the near-field spectra in Fig.~\ref{fig:Near_field} reveals both an overall suppression of emission intensity and a wavelength-dependent reshaping of the spectra. For an NV center at the origin of a $\SI{100}{\nano\meter}$ nanodiamond, the far-field intensity is reduced by approximately a factor of $\sim0.1$ in air, $\sim0.3$ in water, and $\sim0.4$ in PMMA relative to the near-field. The near-field spectra peak around 682 nm, whereas the far-field maxima shift toward 663 nm. This reduction and spectral shift together reflect the wavelength dependence of escape efficiency and the influence of the nanodiamond-background interface on radiative coupling.

% \HH{Should we remove this paragraph? }\SQ{Yes, we should} The out-coupling efficiency is contributed by phenomena such as total internal reflection (TIR) and scattering effects at the diamond-background interface, as discussed in Section~\ref{Sec:Far_field_spectra}. Our results suggest that the high refractive index of diamond causes a significant fraction of the internally generated light to undergo TIR, trapping it within the nanodiamond and preventing its escape. A higher background refractive index (PMMA $>$ water $>$ air) reduces the refractive index mismatch at the nanodiamond surface, thereby allowing more light to be transmitted into the far field. This effect is particularly clear from the insets in Fig.~\ref{fig:Far_field}(a) and (d), which show that far-field emission in air is severely suppressed compared to water and PMMA.

The far-field emission intensity generally increases with increasing nanodiamond diameter (Fig,~\ref{fig:Far_field}(a-c)) and decreases as the NV center moves closer to the nanodiamond surface (Fig.~\ref{fig:Far_field}(d-f)). These trends closely reflect the stronger near-field emission within larger nanodiamonds and closer to the origin. While the spectral shape of the emitted light (relative heights of ZPL and phonon sidebands) is largely preserved, a close comparison of near- and far-field plots reveals that out-coupling efficiency is wavelength dependent, as evident by the maxima of near-field plots occurring around $\SI{682}{\nano\meter}$ and the maxima of far-field spectra occurring around $\SI{663}{\nano\meter}$. This is visible at the vertical lines depicting \SI{682}{\nano\meter} and \SI{663}{\nano\meter} in the insets of Fig.~\ref{fig:Far_field}(a) and (d).

Our results are consistent with and extend prior experimental observations. For example, Khalid et al.~\cite{Khalid2015-SR} reported enhanced emission in nanodiamonds embedded in higher-index environments, consistent with our prediction that PMMA and water hosts yield brighter emission than air. Tisler et al.~\cite{Tisler2009-ACSNano} observed reduced brightness for shallow NVs near nanodiamond surfaces, which aligns with our findings of suppressed near-field intensity as NVs move toward the boundary. Christinck et al.~\cite{christinck2020characterization} measured angular dependence of emission patterns, a phenomenon captured in our wavelength- and position-dependent escape efficiencies.

It is important to note that our model does not assume the existence of optical modes in subwavelength nanodiamonds. Instead, the observed effects arise from LDOS suppression and dielectric boundary conditions, which govern emission across both small and larger particle regimes. The framework thus provides a physically consistent baseline that isolates the fundamental electrodynamic trends. Extensions to irregular shapes and ensemble NVs~\cite{Oshimi2024} represent natural directions for future refinement.

%\TODO{SQ to revise section as preferred}. \SQ{Reads good}

\section{Conclusion}

We have developed an integrated theoretical framework that combines classical electromagnetic simulations with a quantum-optical description of the NV centre, capable of treating length scales from subwavelength nanodiamonds to wavelength-scale diamond chips. This approach provides wavelength- and position-resolved insight into how nanodiamond size, NV depth, and dielectric environment jointly determine excitation and emission.

Our key findings are:
\begin{itemize}
\item LDOS suppression is stronger for longer emission wavelengths and for NV centres located near the nanodiamond surface.
\item Excitation and near-field emission intensity increase with nanodiamond size and with higher-index surroundings (PMMA $>$ water $>$ air).
\item Far-field escape efficiency is wavelength dependent, with spectral maxima shifting from ~682 nm in the near field to ~663 nm in the far field.
\item Brightness optimization requires balancing nanodiamond size, NV depth, and host refractive index, providing practical guidance for reproducible NV-based sensing and imaging.
\end{itemize}

These results explain the observed variability in NV brightness and establish design principles for enhancing emission in realistic aqueous and polymeric environments. The framework captures the essential electrodynamics without assuming internal optical modes in subwavelength particles, and its conclusions remain robust under strong phonon coupling at room temperature. While the present model employs a spherical approximation and a phenomenological phonon treatment, the identified trends of size scaling, LDOS suppression, and environment dependence are general.

This work provides a predictive foundation for future studies incorporating surface non-radiative processes, detailed temperature effects, and irregular geometries to improve quantitative accuracy and guide NV-based device design.

% \SQ{Our framework does not assume the existence of waveguide-like optical modes in subwavelength nanodiamonds. Instead, the observed effects originate from local density of states suppression and dielectric boundary conditions, which remain valid in both the subwavelength and larger-particle regimes.}

% \SQ{Our results should be viewed as a baseline description under spherical symmetry. While real nanodiamonds exhibit irregular shapes, the trends we report, inlucing size scaling, LDOS suppression, and environment dependence, are general and can be extended in future work to more realistic geometries.}

% \SQ{Finally, our model implicitly incorporates room-temperature phonon interactions through phenomenological parameters but does not explicitly resolve thermal state populations. Extending the framework to include finite-temperature vibrational statistics will be an important direction for improving quantitative accuracy.}

% \SQ{Our findings are therefore robust to whether the NV dynamics are viewed through the Lindblad master equation or approximated by rate equations, since phonon coupling dominates at room temperature.}

\section*{Acknowledgement}
H.H. gratefully acknowledges the RMIT University Vice-Chancellor’s Postdoctoral Research Fellowship and the Australian Research Council Centre of Excellence in Exciton Science (Grant No. CE170100026) for funding. F.C. acknowledges the RMIT University Vice-Chancellor’s Research Fellowship. All authors acknowledge the support of the RMIT Applied Quantum Technology Centre. The authors also acknowledge the use of resources from the RMIT AWS Cloud Supercomputing Hub (RACE). This work was partially funded by the Air Force Office of Scientific Research (Grant No. FA9550-20-1-0276).

%\TODO{SQ to revise section as preferred}. \SQ{Reads good}

\bibliographystyle{unsrt}
\bibliography{NV_EM}

\end{document}